\documentclass[preprint,amsmath,superscriptaddress,prl]{revtex4}  
\usepackage{hyperref} 
\usepackage[dvips]{graphicx}
\usepackage{color}
\usepackage{amsfonts}
\usepackage{subcaption}
\usepackage{soul}
\usepackage{amsthm}
\usepackage{comment}

\theoremstyle{definition}

\theoremstyle{plain}

\begin{document} 
\title{
Universal 
finite-time scaling
in the 
transcritical,
saddle-node, 
and pitchfork discrete and continuous
bifurcations
}
\author{
\'Alvaro Corral}
\affiliation{%
Centre de Recerca Matem\`atica,
Edifici C, Campus Bellaterra,
E-08193 Barcelona, Spain
}\affiliation{Departament de Matem\`atiques,
Facultat de Ci\`encies,
Universitat Aut\`onoma de Barcelona,
E-08193 Barcelona, Spain
}\affiliation{Complexity Science Hub Vienna,
Josefst\"adter Stra$\beta$e 39,
1080 Vienna,
Austria}
Corresponding author 
alvaro.corral@uab.es
\begin{abstract}
Bifurcations are one of the most remarkable features of dynamical systems.
Corral et al. [Sci. Rep. 8(11783), 2018] showed the existence of scaling laws describing the
transient (finite-time) dynamics in discrete dynamical systems close to a bifurcation point,
following an approach that was valid for the transcritical as well as for the saddle-node bifurcations.
We reformulate those previous results 
and extend them 
to other discrete and continuous bifurcations, remarkably the pitchfork bifurcation.
In contrast to the previous work, 
we obtain a finite-time bifurcation diagram directly from the scaling law, 
without a necessary knowledge of the stable fixed point.
The derived scaling laws provide a very good and universal description of the transient behavior of the systems
for long times and close to the bifurcation points.
%
\end{abstract} 

\date{\today}

\maketitle




\section{Introduction}


Bifurcations involve a sudden and qualitative change in
the asymptotic solutions of a dynamical system when one parameter (or more) is gradually modified \cite{Strogatz_book}.
This implies that
bifurcations are able to explain profound variations in the stationary state
of many real systems
when some conditions (given by the bifurcation parameter)
are slightly different.
These real systems include
physical, chemical, biological, environmental, 
and any sort of system whose evolution can be formulated in terms of ordinary differential equations 
(or their discrete counterpart).
Remarkable examples are lasers, oscillating chemical reactions, activity of genes, animal or plant populations, and epidemic spreading.
\cite{Strogatz_book,Sole_phase_transitions}.



Trying to strength the connection between branching processes
\cite{Harris_original,branching_biology,Corral_FontClos}
and thermodynamic phase transitions \cite{Stanley,Yeomans1992,Stanley_rmp}, 
as well as to better understand 
the emergence of critical behavior in the infinite system-size limit,
Garcia-Millan et al. \cite{GarciaMillan,Corral_garciamillan}
studied the renowned Galton-Watson process.
They
analytically established the existence of finite-size scaling
\cite{Privman}
and universality \cite{Stanley_rmp}
in the relation between the probability of survival 
at different finite times
and the mean offspring number per individual
when the offspring distribution is arbitrary 
(with the only requirement of a finite variance;
the opposite case has been studied more recently 
and leads to a continuum of universality classes
\cite{Falco_Corral}).
Thus, the Galton-Watson process displays a sudden change 
analogous to a phase transition.
Notably, the aforementioned authors 
\cite{GarciaMillan,Corral_garciamillan} found
an exact analytical expression for 
the scaling function appearing in the scaling law 
describing 
the probability of survival.
Other aspects of the finite-size properties 
of the Galton-Watson process
were analyzed in Ref. \cite{Corral_garcia_moloney_font}.

The same authors 
\cite{GarciaMillan,Corral_garciamillan}, 
however, did not pay attention to the fact that there is a transcritical bifurcation underlying the phase transition in the Galton-Watson process,
as the time evolution of the probability of survival is governed by a discrete dynamical system.
Reference \cite{Corral_Alseda_Sardanyes} put the results of 
Garcia-Millan et al. \cite{GarciaMillan,Corral_garciamillan}
in the context of bifurcations
in dynamical systems, 
showing that the results were valid beyond the Galton-Watson process, 
as they arose from a universal, previously unnoticed
scaling characteristic of the transcritical bifurcation.
It was also shown \cite{Corral_Alseda_Sardanyes}
that this property was present in the saddle-node bifurcation as well.
It is noteworthy how, in the approach of Ref. \cite{Corral_Alseda_Sardanyes}, 
finite-size scaling in the Galton-Watson process
translates into 
finite-time scaling in the framework of dynamical systems
(understanding the independent variable of the system as time,
yielding a description of the transient dynamics),
due to the fact that
in the former process, 
system size (given by the depth of the underlying branching tree) 
and
time (given by the number of generations) 
are so intertwined that they are in fact the same thing
(although the evolution of dynamical systems is more naturally described in terms of time).

In the present article we are guided by this parallelism 
between 
phase transitions
and
bifurcations in dynamical systems, summarized in Table \ref{table1}.
We provide a clear case in which statistical-physics concepts can be used to better understand dynamical systems, concretely, bifurcations.
Despite of the fact that
our approach looks similar to that in Ref. \cite{Corral_Alseda_Sardanyes}, 
we expand the results of that work in several directions.
In Sec. 2 we briefly summarize the results of Ref. \cite{Corral_Alseda_Sardanyes}.
Section 3 deals with
discrete dynamical systems, 
for which we derive and test a scaling law for the distance of the solution to a fixed point that can be stable or unstable, and when this is constant (independent on the bifurcation parameter), 
we directly obtain a ``finite-time bifurcation diagram'' 
that is valid for a variety of bifurcations, including the pitchfork one 
(in addition to the transcritical and saddle-node cases studied in Ref. \cite{Corral_Alseda_Sardanyes}).
In Sec. 
4 we extend the previous results \cite{Corral_Alseda_Sardanyes} as well as our new results to continuous dynamical systems, whereas in the subsequent section we study an alternative scaling law for the distance of the solution to a stable fixed point, exploring also the limits of validity of these scaling laws.
Finally, after discussing the results,
an Appendix shows how the well-known phenomenon of critical slowing down is contained in our scaling laws, 
allowing the derivation of the exponents of the temporal power-law decay to the attractor at the bifurcation point 
as well as the values of the characteristic decay times outside this point.




\begin{table}[h]
\begin{center}
\caption{\label{table1} 
Parallelisms between thermodynamic phase transitions, branching processes, 
and bifurcations in dynamical systems
(thermodynamic phase transitions are illustrated with the example of a magnetic material
at zero external field).
The notation we use in the main text is introduced in the third column.
In dynamical systems, fixed points are sometimes referred to as critical points, 
but we do not use such terminology 
as there is no correspondence with the usage of the same term in phase transitions.
}
\smallskip
\begin{tabular}{|l|l|l|}
\hline
Phase transitions& Branching processes &Dynamical systems\\
\hline\hline
Control parameter & Mean offspring number  & Bifurcation parameter \\
(as temperature $T$) &$\langle N\rangle$& $\mu$\\
\hline
Order parameter &Probability of survival (at infinity)& Fixed-point attractor \\ 
in thermodynamic limit& $\mathcal{P}$& (or stable fixed point)\\
(as magnetization $\mathbb{M}$) && $q$\\
\hline
Critical point & Critical point & Bifurcation point \\
(as $(T^*, \mathbb{M}^*)$ with $\mathbb{M}^*=0$)
 &
$(\langle N\rangle^*,\mathcal{P}^*)$ with $\langle N\rangle^*=1$, $\mathcal{P}^*=0$
& $(\mu^*$, $q^*)$\\
\hline
System size &Number of generations & Independent variable \\
 && (time $n$, $\ell$, $t$, $\tau$)\\
 \hline
 Finite-size scaling&Finite-size scaling $=$&Finite-time scaling\\
 &finite-time scaling&\\
\hline
\end{tabular}
\par
\end{center}
\end{table}

\section{Previous research: 
The discrete transcritical and saddle-node bifurcations} 

Let us consider the following one-dimensional discrete dynamical system or map
\cite{Strogatz_book}
\begin{equation}
x_{n+1}=f(x_n)
\label{discreta00}
\end{equation}
where $n=0,1\dots$ 
denotes
a discrete time (counting time generations)
and $x_n$ are real numbers; thus, $f(x)$ (also called a map) 
is a real function.
It is assumed that the system has 
a linearly stable fixed point 
(also referred to here as an attractive fixed point) at
$x=q$, with $q=f(q)$ and $|f'(q)|<1$
(the prime denotes differentiation and
the bars absolute value).
It is implicit that the function $f(x)$ depends on a 
bifurcation parameter $\mu$ and so, $q$ 
may depend on $\mu$ as well.

Reference \cite{Corral_Alseda_Sardanyes}
(based on results from Ref. \cite{GarciaMillan})
showed that 
in such a discrete dynamical system,
for large values $\ell$ of the discrete time,
close to a bifurcation point,
and when the initial condition $x_0$ belongs to the basin of attraction of the fixed point,
the difference between 
the stable fixed point $q$
and
the solution at time $\ell$
fulfills
a finite-time scaling law, 
\begin{equation}
q-x_\ell=\frac 1 {D^*\ell}\, G(z),
\label{llei0}
\end{equation}
where the rescaled variable $z$ is defined as 
\begin{equation}
z=\ell(\tilde m-1),
\label{zlm1}
\end{equation}
with $\tilde m=f'(q)$, and
$D^*=f''(q^*)/2\ne 0$,
with $q^*$ denoting the value that the fixed point takes at the bifurcation point.
Remarkably, the expression does not depend on the initial condition.
Even more remarkably, the scaling function $G(z)$ 
turned out to be independent on $f(x)$ and therefore ``universal,'' being
\begin{equation}
G(z)=\frac{ze^z}{e^z-1}.
\label{scalingfunc}
\end{equation}
It is worth mentioning that Eq. (\ref{llei0}),
together with Eq. (\ref{zlm1})
constitutes a clear example of what is called a scaling law
in statistical physics \cite{Christensen_Moloney,Corral_csf}.
Other authors use the term scaling law to refer to (univariate) power-law functions;
certainly, power laws are a particular case of scaling laws 
(when the scaling function is constant);
nevertheless, in this article a scaling law refers to something akin to Eqs. (\ref{llei0}) plus (\ref{zlm1}) and not to a simple power law.

Note that, as the fixed point 
$q$ is linearly stable,
then
$\tilde m<1$, 
and $z<0$. 
Restricted to $z\le 0$ (extending $z$ to the bifurcation point $z=0$, 
where the attractor loses its linear stability) the function $G(z)$
is monotonically increasing with $z$, with
a maximum at $z=0$.
This indicates that the closer to the bifurcation point, 
the longer it takes to reach a small neighborhood of the attractor;
this is quantified by the scaling law,
and the rescaled variable $z$ provides the appropriate way  
to compute the ``distance'' to the bifurcation point
(depending on the value of $\ell$).

In consequence, 
the attractors of certain one-dimensional discrete dynamical systems 
given by Eq. (\ref{discreta00})
are approached in the same universal way,
with the only condition that $f''(q^*)\ne 0$.
This was found to be valid for the transcritical and the saddle-node bifurcations \cite{Corral_Alseda_Sardanyes}.
Knowing the dependence of $q$ on the bifurcation parameter, 
it was also possible to find scaling laws for the relation between 
$x_\ell$ and $\tilde m$ 
(eliminating $q$ from Eq. (\ref{llei0}))
in the transcritical and saddle-node bifurcations
(for large $\ell$ and close to the bifurcation point).
Nevertheless, this second scaling laws were not fully universal, 
in the sense that they were different for each bifurcation
(although the scaling functions for each one were mathematically
related through simple transformations).

\section{
Finite-time bifurcation diagram
in discrete bifurcations}


\subsection{Theoretical derivation}

We start with the one-dimensional discrete dynamical system
considered in Ref. \cite{Corral_Alseda_Sardanyes} and given by Eq. (\ref{discreta00}).
An important difference between our approach and 
that previous work 
is that, for us, 
the reference fixed point can be 
either linearly stable or linearly unstable (or neutral),
and then, instead of by $q$, it will be denoted by $p$, verifying
$|f'(p)| < 1$ or $|f'(p)| > 1$ (or $|f'(p)| = 1$), 
respectively. 
In other words, the stability of the fixed point does not matter for us.
Also, the previous work \cite{Corral_Alseda_Sardanyes} assumed,
as mentioned,
$D^*=f''(q^*)/2\ne 0$,
which holds for the transcritical and saddle-node bifurcations,
but not in other cases.

We assume here that $f''(p^*)$ 
and $f''(p)$ 
may be zero
and define $P=-f^{(k)}(p)/k!$,
with $k\ge 2$ the lowest 
order of the derivative 
for which 
$f^{(k)}(p)\ne 0$,
and with $k$ in parenthesis denoting the $k-$th order derivative
(with $k$ integer;
the case studied in Ref. \cite{Corral_Alseda_Sardanyes} corresponds to $k=2$).
%
%
Given the initial condition $x_0$,
the $n-$th iteration of $f$ is 
$f(\dots f(f(x_0))\dots)=f^n(x_0)$.
For sufficiently large $n$, we perform a Taylor expansion 
of $f(f^n(x_0))$ around 
the (stable or unstable) fixed point $p$, so
\begin{equation}
    f^{n+1}(x_0)= p+m(f^n(x_0)-p)-P(f^n(x_0)-p)^k
+ \mathcal{O}(f^n(x_0)-p)^{k+1},
\label{lacinco}
\end{equation}
with $m=f'(p)$.
If $p$ is attractive, as $n$ is large,
it is obvious that $f^n(x_0)$ will be close to $p$;
however,
when $p$ is unstable (that is, repulsive), 
we assume that there will be an attractive fixed point $q$ in between $x_0$ and $p$, 
and $f^n(x_0)$ will be close to $q$; 
moreover, close to the bifurcation point, $q$ will be close to $p$ and therefore $f^n(x_0)$ will be close to $p$.
It is implicit then that the bifurcation is continuous.
It is assumed also that $x_0$ belongs to the basins of attraction of the attractive fixed point. 

Introducing the ``distance'' (which can be negative) 
between the solution at time $n$ and the fixed point $p$ as
$d_n=f^n(x_0)-p$ we can write Eq. (\ref{lacinco}) as 
$d_{n+1}=m d_n -P d_n^k$, up to order $k$.
It is convenient to define $c_n=1/d_n^\alpha$, with $\alpha$ arbitrary
\cite{footnote_aina},
and from here, as $d_n$ will be small,
$$
c_{n+1}=\frac {c_n}{m^\alpha}
+ \frac{\alpha P d_n^{k-1-\alpha}}{m^{\alpha+1}}
$$
plus higher-order terms \cite{Aina_report}.
The expression is particularly simple taking $\alpha=k-1$, 
then, iterating
$\ell$ times we arrive at
$$
c_{n+\ell} \simeq \frac{c_n}{m^{(k-1)\ell}} +
\frac {(k-1)P}{m^k}\left(1+\frac 1 {m^{k-1}}+\dots +\frac 1 {m^{(k-1)(\ell-1)}}\right)
$$$$
= \frac{c_n}{m^{(k-1)\ell}} +
\frac {(k-1)P}{m^k} \frac{m^{(k-1)\ell}-1}{m^{(k-1)(\ell-1)}(m^{k-1}-1)}.
$$
%
We deal with the existence of a bifurcation at $f'(p)=1$
(ignoring bifurcations given by $f'(p)=-1$);
then, similarly as in Eq. (\ref{zlm1}), \cite{Corral_Alseda_Sardanyes}, we define, 
\begin{equation}
y=\ell(m-1),
\label{yellmm1}
\end{equation}
with $m=f'(p)$ and therefore
$m^{\ell}\rightarrow e^{y}$
in the limit of large $\ell$. 
From here,
$$
c_{n+\ell}\simeq\frac{c_n}{e^{(k-1)y}} 
+ (k-1)P \, \frac{e^{(k-1)y}-1}{e^{(k-1)y}(k-1)y/\ell}.
$$
We may define $\ell'=n+\ell$ but for $\ell \gg n$ we have $\ell\simeq \ell'$.
In that case, the second term dominates (it is proportional to $\ell$), and
$$
c_{\ell }\simeq\frac{(k-1) P^* \ell}{G((k-1)y)},
$$
with $G(y)$ given by Eq. (\ref{scalingfunc})
and approximating (to the lowest order) 
$P\simeq P^*$
(the value at the bifurcation point,
as for large $\ell$ and finite $y$, 
the parameter
$m$ has to be close to one and the system has to be close to the bifurcation).

Coming back to the distance between the solution and the fixed point, 
we have that $d_\ell^{k-1}=1/c_\ell$ and then
\begin{equation}
\left(f^\ell(x_0)-p\right)^{k-1}=
{\frac 1 {(k-1) P^* \ell}\,G((k-1)y)},
\label{leygeneralizada}
\end{equation}
which constitutes,
together with Eq. (\ref{yellmm1}), 
a finite-time scaling law.
From now on
we write it as an equality, 
although it is only 
exact in the limits $\ell \rightarrow \infty$
and $m\rightarrow 1$.
This scaling law contains the one in Ref. \cite{Corral_Alseda_Sardanyes}
as a special case, taking $k=2$ and considering that the fixed point
is stable (changing $p$ for $q$).
%
It is remarkable that we obtain the same scaling function $G(y)$ as in Refs. \cite{GarciaMillan,Corral_Alseda_Sardanyes},
despite our problem is more general.
Note that we use $y$ instead of $z$, as $z$ in Ref. \cite{Corral_Alseda_Sardanyes}
was restricted to $z \le 0$, but as $p$ can now be unstable we do not have such a restriction.
Moreover, in our case, 
$m$ can be considered to play the role of an alternative, ``natural'' bifurcation parameter 
(and therefore has to be related to $\mu$)
as we do not have the restriction $\tilde m < 1$.
This means that $m$ quantifies the distance to the bifurcation
point (given by $m^*=1$).
We also realize that, when $k$ is odd,
$P^*$ has to be larger than zero
(for the scaling law to hold),
as the scaling function $G$ is positive for any finite $y$
but $x_\ell-p$ can take both positive and negative values.
In contrast, when $k$ is even there is no restriction on the sign of $P^*$, 
and $x_\ell - p$ will have the same sign as $P^*$;
therefore,
in this case the scaling law is only valid at one side 
in the bifurcation diagram.
It is straightforward but very useful to realize that the scaling law (\ref{leygeneralizada})
can also be written as
\begin{equation}
f^\ell(x_0)-p=
(\pm 1)^k 
\left({\frac 1 {(k-1) P^* \ell}\,G((k-1)y)}\right)^{1/(k-1)},
\label{leygeneralizada2}
\end{equation}
where the term $(\pm 1)^k$ arises from the real values of $\sqrt[k-1]{1}$.

\subsection{Bifurcation arising from the smoothness of the scaling function}

Using the obtained scaling law,
we can see 
how the sharp change represented by the bifurcation emerges in the limit $\ell \rightarrow \infty$ from the asymptotic properties of a continuous and smooth scaling function $G(y)$.
Indeed, from Eq. (\ref{scalingfunc}),
$$
G(y) \rightarrow \left\{
\begin{array}{ll}
0 & \mbox{ for } y\rightarrow -\infty,
\\ 
1 & \mbox{ for }  y\rightarrow 0,
\\
y & \mbox{ for }  y\rightarrow \infty,\\
\end{array}
\right.
$$
and substituting in the scaling law, Eq. (\ref{leygeneralizada2}),
\begin{equation}
f^{\ell}(x_0) -p \rightarrow
\left\{
\begin{array}{lll}
0 
& \mbox{ for } m<1
& \mbox{ and } \ell \rightarrow \infty,\\ 
(\pm 1)^k
\sqrt[k-1]{\frac{1}{(k-1)P^*\ell}} 
& \mbox{ for } m=1,
&
\\
(\pm 1)^k
\sqrt[k-1]{\frac{m-1}{P^*}} 
& \mbox{ for } m>1
& \mbox{ and } \ell \rightarrow \infty,\\
\end{array}
\right.
\label{ladelallave}
\end{equation}
with the obvious meaning that
when $m<1$, the fixed point $p$ is (linearly) stable and the
solution tends to it for $\ell \rightarrow\infty$,
whereas when $m>1$, the fixed point $p$ is unstable and
the solution tends to another fixed point $q$.

In the latter case,
the value of the stable fixed point $q$ (minus $p$) is obtained
from the Taylor expansion around the unstable fixed point $p$,
without any knowledge of $q$. 
A case of great interest is when $p$ is zero, or constant 
(for all the considered values of $\mu$),
for which the scaling law yields the equivalent of the bifurcation diagram but for finite times,
without the need of calculating the stable fixed point $q$;
i.e., only the knowledge of the constant (stable or unstable) fixed point $p$ is required.

In the previous subsection we have seen that, when $k$ is odd, 
then $P^*>0$ and $x_\ell-p$ can take any sign.
Now we can see, from Eq. (\ref{ladelallave}),
which yields $q-p$ (with $q=p$ for $m<1$),
that $x_\ell > q$ when $q>p$
and $x_\ell < q$ when $q<p$.
The reason is that $G(y)>y$ when $y>0$
and $G(y)>0$ when $y<0$ (in fact, always).
Thus, given an attractive fixed point $q$, the scaling law
is only valid for initial conditions at one of its sides.
The same is true when $k$ is even
(but this is already clear from the previous subsection).


\subsection{Numeric verification} 

For a given dynamical system, characterized by a value of $k$,
the fulfillment of the scaling law 
(\ref{leygeneralizada2}) 
for any value of the discrete time $\ell$ (large) 
and $m$ (close to one, related to the bifurcation parameter)
must become apparent 
displaying $\sqrt[k-1]{(k-1)P^* \ell\,}\,(x_\ell-p)$
versus $y=\ell(m-1)$,
yielding the shape of the function 
$(\pm 1)^k [G((k-1)y)]^{1/(k-1)}$,
which does not depend neither on $\ell$ nor on $m$ individually, 
but on $\ell(m-1)=y$
(note the metric factor $k-1$ 
multiplying $y$ in $G()$ \cite{Privman}).
We obtain in this way a law of corresponding states \cite{Stanley}
that signals the existence of universality, 
as only $k$ and the first non-zero derivatives $m$ and $P^*$ of the map
$f(x)$
matter.
%
However, there is an additional level of universality,
as the same scaling law
is valid for different bifurcations, 
taking into account that the value of $k$ may change
from one bifurcation to another.
In that case, for different $k$, 
Eq. (\ref{leygeneralizada}) indicates that
one should display the more general rescaling
${(k-1) P^* \ell \, (x_\ell-p)^{k-1}}$
versus
$(k-1)\ell(m-1)$,
leading to a collapse of all the data into the unique function $G$.



\begin{figure}[h]
   \includegraphics[width=0.5\linewidth]{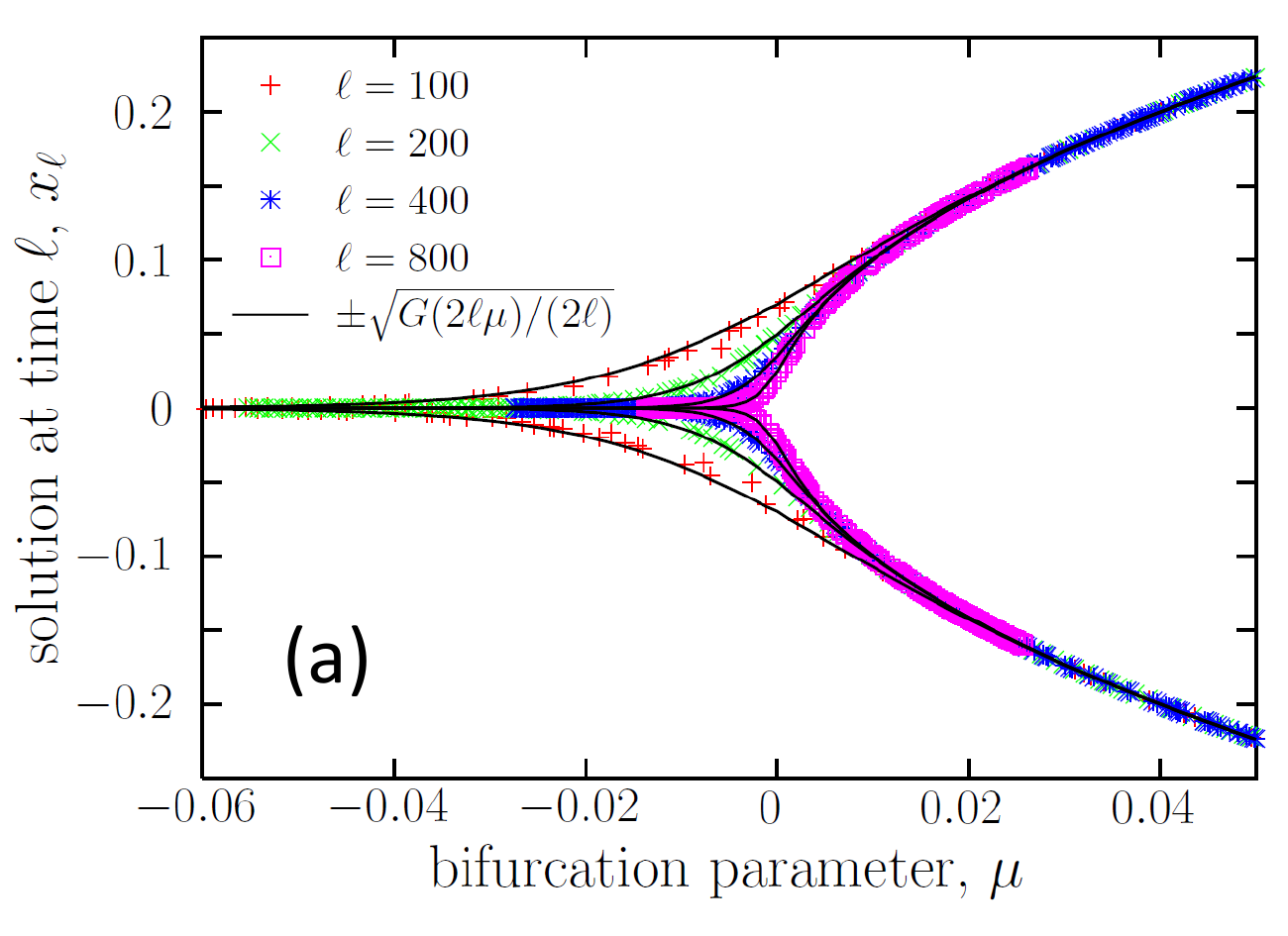}
    \includegraphics[width=.5\linewidth]{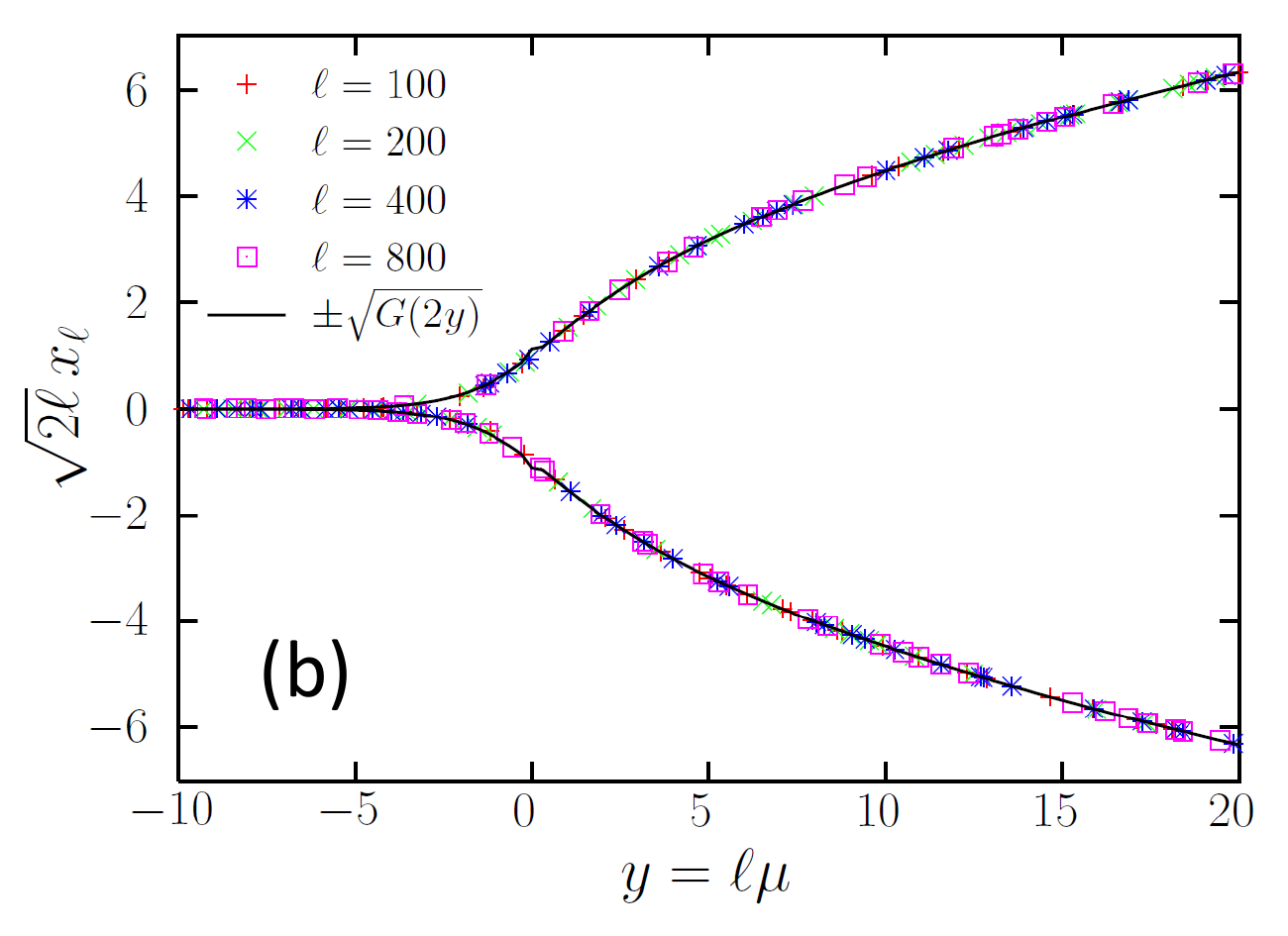}
    \caption{
(a) $x_\ell$ versus $\mu$ at different values of the time $\ell$ 
for the discrete dynamical system given by
Eq. (\ref{fxk}) with $k=3$,
yielding the normal form of the discrete (supercritical) pitchfork bifurcation.
Different initial conditions are taken, uniformly distributed 
between $q+0.1$ and $q+0.6$ for $q \ge 0$
and between
$q-0.6$ and $q-0.1$ for $q\le 0$.
(b) A part of the data (for clarity sake) with rescaled axes, 
representing
$\sqrt{2\ell} \,x_\ell$ versus $y=\ell \mu$, 
as derived from Eq. (\ref{leygeneralizada2}).
The function $\sqrt{G(2y)}$ becomes apparent with the rescaling,
and provides the ``microscopic'' smooth form of the pitchfork bifurcation
(in contrast to the usual well-known, sharp ``macroscopic'' picture).
}
    \label{pitch}
\end{figure}

\begin{figure}[h]
\includegraphics[width=.5\linewidth]{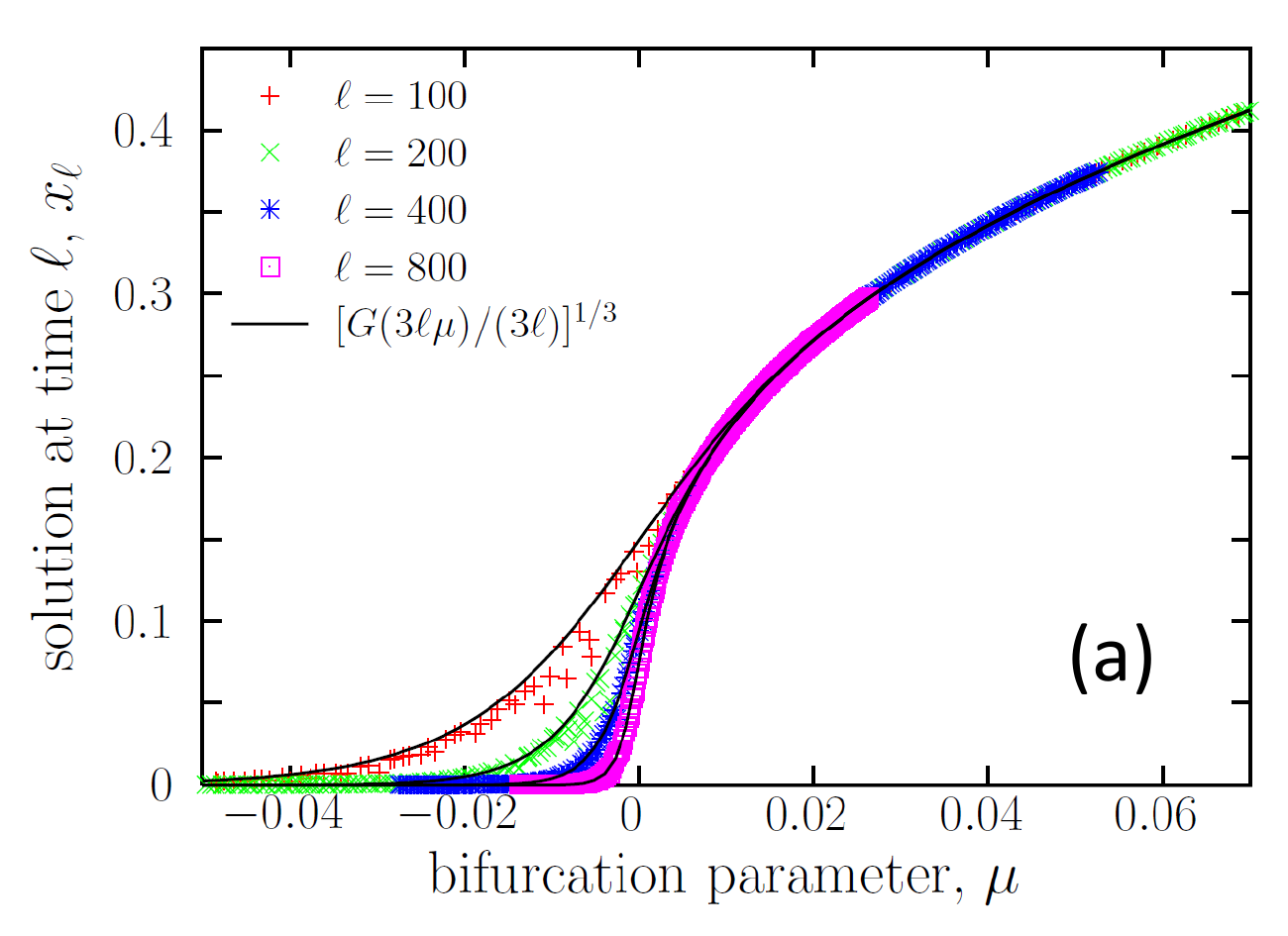}
\includegraphics[width=.5\linewidth]{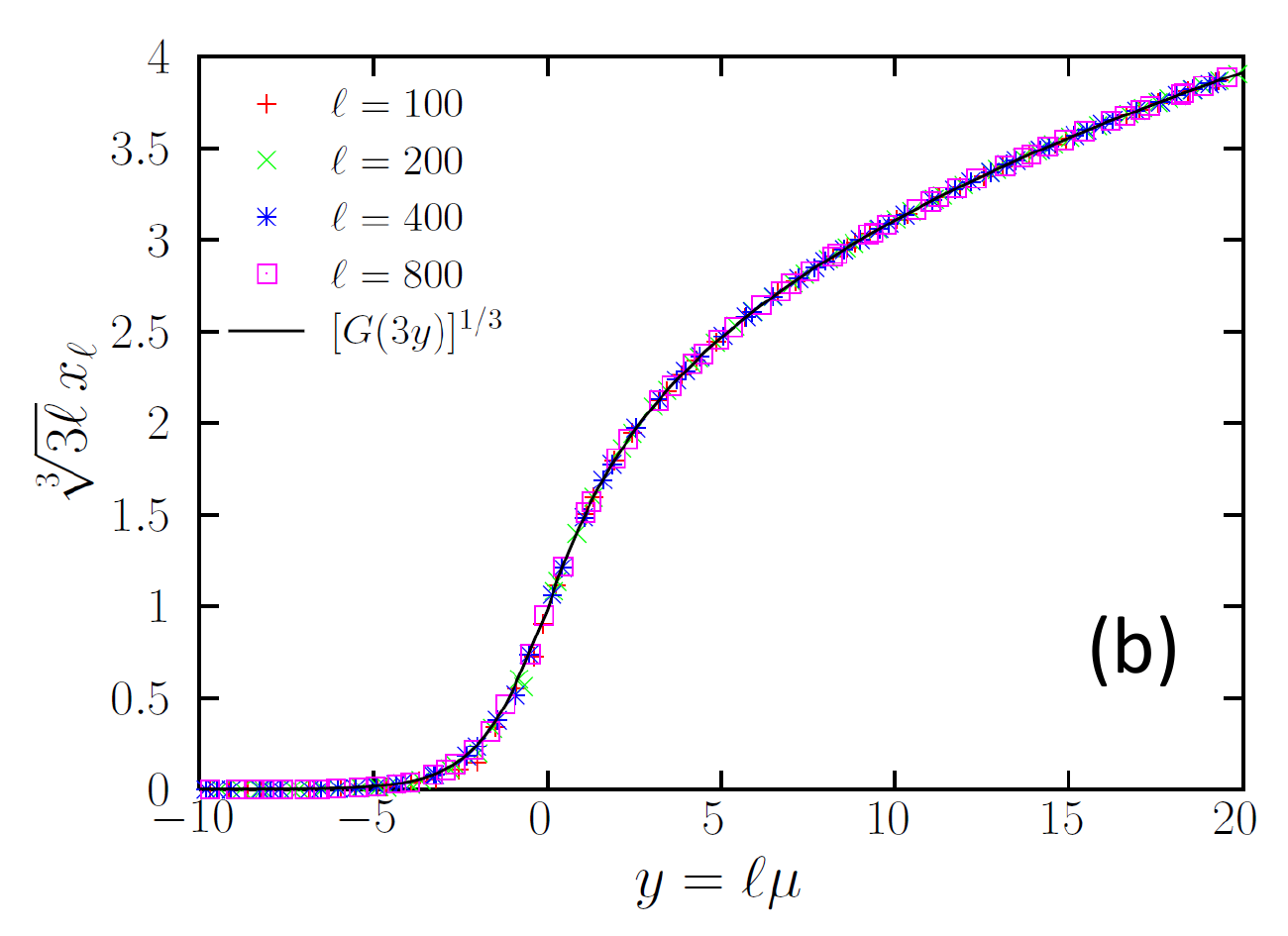}\\
\includegraphics[width=.5\linewidth]{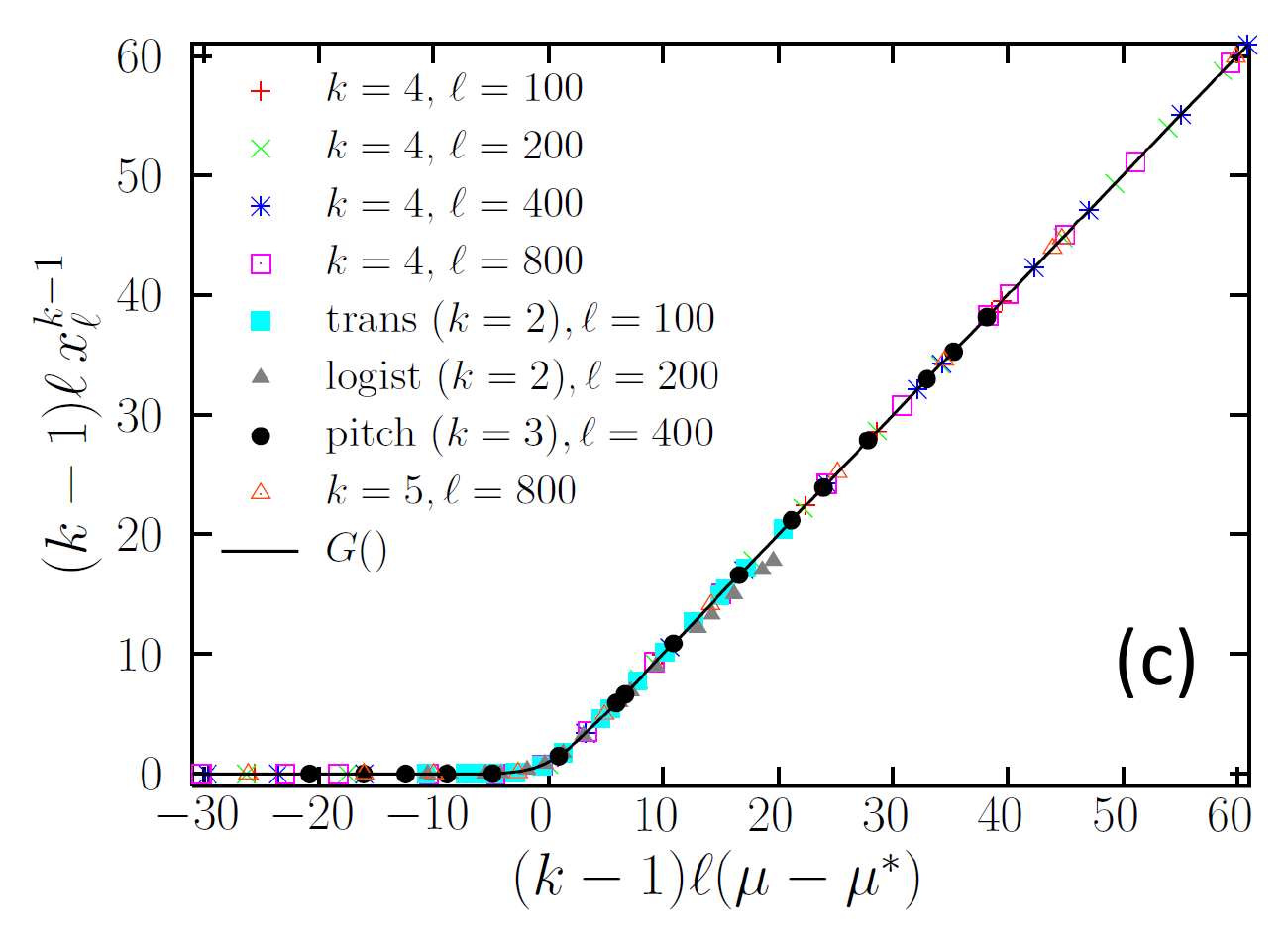}
    \caption{
(a) $x_\ell$ versus $\mu$ at different values of the time $\ell$ 
for the discrete dynamical system 
given by Eq. (\ref{fxk}) with $k=4$.
Initial conditions taken in the same way as in previous figure.
(b) A part of the data (for clarity sake) with rescaled axes, 
representing
$\sqrt[3]{3\ell} \,x_\ell$ versus $y=\ell \mu$,
yielding the function $\sqrt[3]{G(3y)}$.
(c) Same data with the more general rescaling
obtained from Eq. (\ref{leygeneralizada}), 
displaying
${(k-1)\ell} \,x_\ell^{k-1}$ versus $(k-1)y=(k-1)\ell (\mu-\mu^*)$,
together with other maps, 
given by
$k=2$ (transcritical), $k=3$ (pitchfork, previous figure), and $k=5$,
and $f(x)=\mu x (1-x)$ (logistic, corresponding to $k=2$).
Notice that, in some sense, we are transforming all bifurcations diagrams with $k\ge 3$
into the one of the transcritical bifurcation 
(for which $k=2$ and then the exponent $k-1$ does not introduce any changes to it).
%
%
$\mu^*=1$ for the logistic and $\mu^*=0$ for the rest.
}
    \label{k4}
\end{figure}

We can test the validity of the scaling law with specific examples.
Figures \ref{pitch} and \ref{k4} analyze, close to the bifurcation point,
the behavior of several discrete dynamical systems given by 
\begin{equation}
f(x)=(1+\mu)x-x^k,
\label{fxk} 
\end{equation}
where the bifurcation parameter is $\mu$
and $k=2,3...$ corresponds to different bifurcations.
In this way, 
$k=2$ yields the normal form of the transcritical bifurcation
and
$k=3$ gives the normal form of the (so-called supercritical) pitchfork bifurcation \cite{Strogatz_book}.
Independently of the value of $k$,
all the dynamical systems 
represented by this $f(x)$
have $p=0$ as a fixed point 
(it does not matter if stable or unstable)
for all $\mu$.
Then, $m=f'(p)=1+\mu$ 
(thus, all the bifurcations for different values of $k$ take place at $\mu=0$), 
$y=\ell (m-1)=\mu\ell$ and $P=-f^{(k)}(p)/k!=1$.
Figures \ref{pitch}(a) and \ref{k4}(a) display,
for $k=3$ and $4$ respectively,
the solution $f^{\ell}(x_0)$ as a function of the bifurcation parameter $\mu$ for different values of $\ell$ (and changing also $x_0$). 
What we obtain are the usual profiles that approach the underlying bifurcation diagram in the limit 
$\ell \rightarrow \infty$.

Figures \ref{pitch}(b) and \ref{k4}(b) display essentially 
the same data 
but applying the rescaling given by the scaling law, 
under the form directly suggested by
Eq. (\ref{leygeneralizada2}).
The observed nearly-perfect data collapses 
clearly support the validity of the scaling law.
Figure \ref{k4}(c) 
gives further support, 
including, in addition,
results from $k=2$ to $k=5$,
as well as from the logistic map
at its transcritical bifurcation,
using the more general rescaling arising directly from Eq. (\ref{leygeneralizada}).
This rescaling, with the ``generalized distance'' $(x_\ell-p)^{k-1}$ to the fixed point,
leads to a very good data collapse; 
this indicates a strong universal behavior for different bifurcations.

Note that the main requirement for the scaling law to hold is
that the successive derivatives of $f(x)$ at $x=p$ are zero from order 2 to $k-1$,
independently of the value of $\mu$.
In the considered examples, the fact that $p$ is constant
($p=0$, in particular) 
is what makes that all the derivatives from order 2 to
$k-1$ are zero (together with the independence of $f^{k}(x)$ on $\mu$).
In contrast, when using the stable fixed point $q$, then
$f''(q^*)=0$ but $f''(q)\ne 0$ if $q \ne q^*$ for $k\ge 3$, 
and then Eq. (\ref{lacinco}) and all the subsequent derivations do not apply.
Therefore, the use of a fixed point $p$ that can be stable or unstable is an important 
improvement with respect the approach of Ref. \cite{Corral_Alseda_Sardanyes}.

\section{Finite-time scaling law for continuous dynamical systems
and unification with the discrete case} 
\label{section_cont}


\subsection{Theoretical derivation}

In this section we 
consider one-dimensional continuous dynamical systems
given by 
\begin{equation}
x'(t)=g(x),
\label{continuous_dynsyst} 
\end{equation}
where 
$x$ is real,
$g(x)$ is a real function, 
and
$t$ denotes continuous time.
%
As in the discrete case,
it is assumed that the system has a (stable or unstable) fixed point at $x=p$, 
with 
$g(p)=0$ (and $g'(p)<0$ or $g'(p)>0$, respectively,
with a bifurcation taking place at $g'(p)=0$)
\cite{Strogatz_book}, 
and that
the initial condition, $x(0)$, belongs to the basin of attraction of the fixed point.
Also, it is implicit that the function $g(x)$ depends on a 
bifurcation parameter $\mu$ whose change 
may change the value of $p$.


In order to study the transient behavior of the continuous dynamical system,
%
%
we may 
approximate this system by a discrete one, 
writing
Eq. (\ref{continuous_dynsyst})
as
$x(t+\Delta)\simeq x(t)+g(x(t)) \Delta$
with $\Delta>0$ sufficiently small.
Therefore, we consider the discrete system given by Eq. (\ref{discreta00}),
$x_{n+1}=f(x_{n})$,
with $n$ the discrete time 
and
\begin{equation}
   f(x)=x + \Delta g(x).
   \label{f_definicio}
\end{equation}
Using a one-step Euler approach
\cite{numeric}, 
with time step $\Delta$,
one can show that 
$x_n$ 
is an approximation of the continuous system's solution,
in the sense of
$x_n  \simeq x(t_n)$ for $t_n=\Delta n$,
taking into account that the initial condition 
has to fulfill $x_0 = x(0)$.


%

Besides, 
it is straightforward to see that both systems have the same fixed points, 
and, due to the fact that $\Delta$ is small enough, 
the stability (and the instability) of the discrete system is the same as that of the continuous one.
Indeed, if $g(p)=0$ with $g'(p)<0$, 
then $f(p)=p$ and, for $\Delta$ small, $|f'(p)| =1- \Delta|g'(p)|<1$
(but still close to one, and thus we avoid the case $f'(p)\simeq -1$),
which are the conditions for existence and linear stability of fixed points in discrete dynamical systems
\cite{Strogatz_book}
(on the other side, if $g'(p)>0$, then $f'(p)>1$).
In addition, $g'(p^*)=0$ implies $f'(p^*)=1$
and therefore, if the continuous system has a bifurcation,
the discrete system has the same bifurcation,
with the same bifurcation point $\mu=\mu^*$. 
In summary, the whole
bifurcation diagram of the continuous system 
is inherited by the discrete system.
In this way,
a continuous dynamical system can be replaced by a discrete one
defined by a close-to-linear map $f(x)$
(in this sense, continuous dynamical systems can be considered a special case of discrete ones). 




Due to the fact that the continuous and discrete dynamical systems 
(\ref{continuous_dynsyst}) and (\ref{discreta00})
are equivalent when the relation (\ref{f_definicio}) holds, 
the continuous one also fulfills the scaling law 
(\ref{leygeneralizada})-(\ref{leygeneralizada2}),
with the same scaling function $G(y)$ given by 
Eq. (\ref{scalingfunc}). 
%
Naturally, for the continuous system
it is more convenient to express the scaling law in terms of the 
variables of the
continuous system.
Thus, 
we substitute 
$f'(p)=1+hg'(p)$
in
$y=\ell(f'(p)-1)$ [Eq. (\ref{yellmm1})]
to get 
\begin{equation}
y=t M=t  g'(p),
\label{ytM}
\end{equation}
with $t=\ell \Delta$ and $M=g'(p)$.
Also,
$P^*\ell=-t g^{(k)}(p^*)/k!=C^* t$,
where we have 
defined $C^*=-g^{(k)}(p^*)/k!$
(then, $M=(m-1)/\Delta$ and $C^*=P^*/\Delta$).
%
%
%
It turns out to be, in the same way as for $f(x)$, 
that $k$ (integer) is the lowest order for which $g^{(k)}(p)\ne 0$,
with $k\ge 2$.

Therefore, 
substituting the previous relations into Eq. (\ref{leygeneralizada2}),
the scaling law for the continuous dynamical system (\ref{continuous_dynsyst}) can be written as, 
\begin{equation}
x(t)-p=
(\pm 1)^k 
\left({\frac 1 {(k-1) C^* t}\,G((k-1)y)}\right)^{1/(k-1)},
\label{llei}
\end{equation}
with
$y$ given by Eq. (\ref{ytM})
and
$G()$ given by Eq. (\ref{scalingfunc}).
Again, the scaling law is valid for large $t$, 
and so for $M=g'(p)$ close to zero (as $y=tM$ is finite), 
and then, if there is a bifurcation,
the scaling law turns out to be valid close to the bifurcation point,
given by $M=M^*= 0$.
%
%

\subsection{Numerical verification in the continuous case}

\begin{figure}[h]
\includegraphics[width=.5\linewidth]{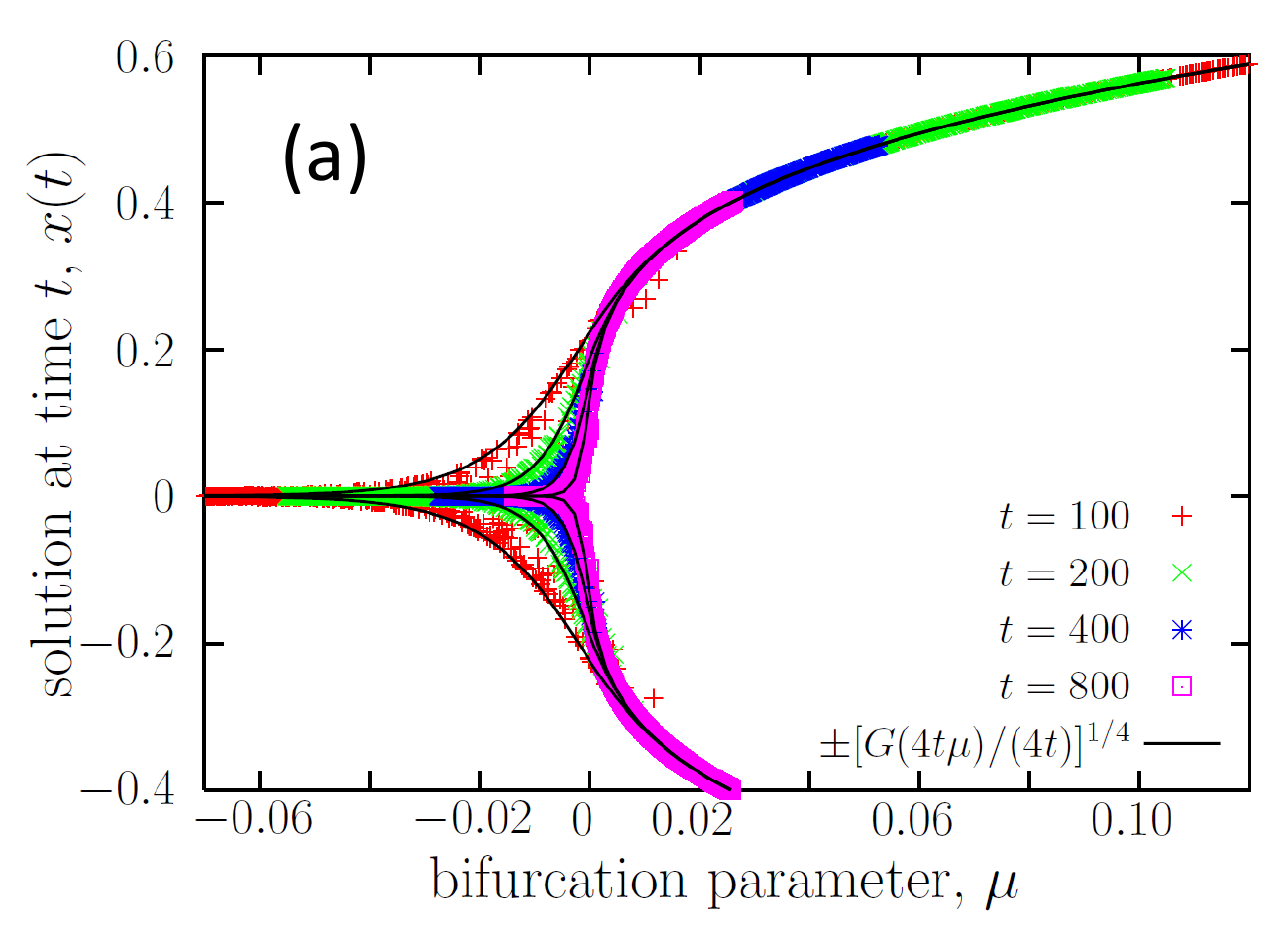}
\includegraphics[width=.5\linewidth]{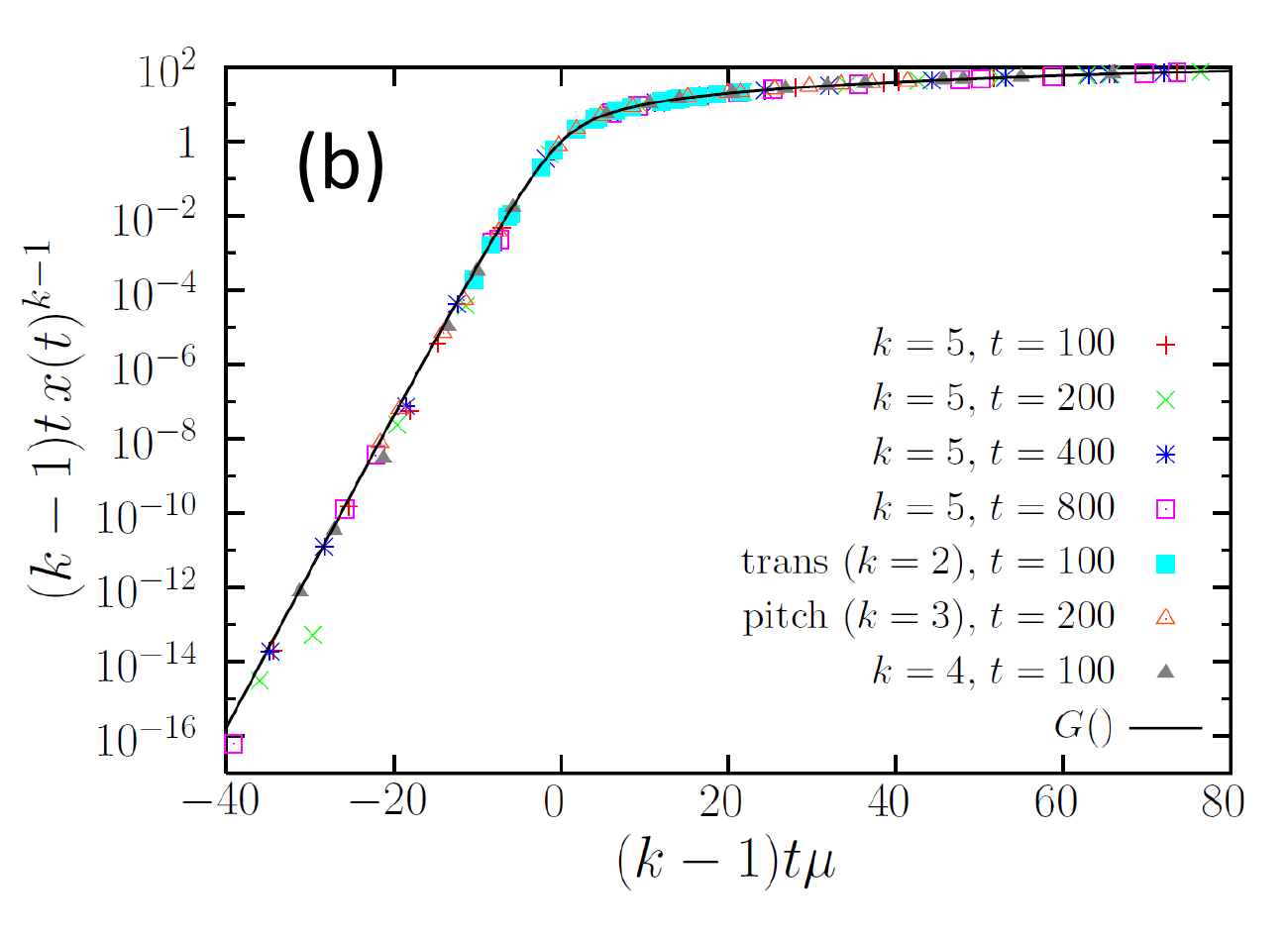}
   \caption{
(a) $x(t)$ versus $\mu$ at different times $t$ 
and using different initial conditions
for the continuous dynamical system
given by $g(x)=\mu x-x^5$ (thus, $k=5$).
Initial conditions are as in the previous figures.
(b) Some part of the data with rescaled axes, 
displaying
${(k-1) t} \,x(t)^{k-1}$ versus $(k-1)y=(k-1) t M$,
together with other continuous systems
with $g(x)=\mu x-x^k$, 
given by
$k=2$ (transcritical), $k=3$ (pitchfork), and $k=4$.
Note that the vertical axis is logarithmic.
}
    \label{Figcont}
\end{figure}

The validity of the finite-time scaling law for continuous bifurcations, Eq. (\ref{llei}), is tested in Fig. \ref{Figcont} using 
the continuous system  given by
$g(x)=\mu x-x^k$ with $k=5$.
Figure \ref{Figcont}(a) 
displays $x(t)$ versus $\mu$
for different values of $t$ and different initial conditions. 
Note that, as in the discrete case,
$p=0$, with $M=g'(p)=\mu$ 
(thus, the bifurcation takes place at $\mu=0$), 
$y=\mu t$, and
$C^*=1$.
Figure \ref{Figcont}(b) shows the results obtained 
when the more general rescaling 
analogous to the one in Eq. (\ref{leygeneralizada})
(and also contained in Eq. (\ref{llei})) is applied,
including other bifurcations given by the $g(x)$ above with different values of $k$.
The vertical axis is logarithmic, 
to test the validity of the scaling law 
when $x(t)$ is very close to zero.
As in the discrete case (previous figures),
the data collapse is nearly perfect, which supports the good description provided by the scaling law (\ref{llei}) 
in the continuous case at finite times.


\subsection{Unification of the discrete and continuous scaling laws}

We can easily unify the scaling laws 
for discrete (\ref{discreta00})
and for continuous (\ref{continuous_dynsyst})
dynamical systems, Eqs. (\ref{leygeneralizada2}) and (\ref{llei}), respectively.
These two equations can be integrated into a single one,
writing it in the form equivalent to Eq. (\ref{leygeneralizada}), which leads to
\begin{equation}
\left(x(\tau)-p\right)^{k-1}=
\frac 1 {(k-1) K^* \tau} \,  G((k-1) y),
\label{lleipm}
\end{equation}
where we have introduced a generic time $\tau$ 
that can be discrete, $\tau=\ell$, or continuous, $\tau=t$,
whereas $x(\tau)$ denotes the solution at time $\tau$ in both cases.
Unifying also the notation 
for the function defining the dynamics 
in Eqs. (\ref{discreta00}) and (\ref{continuous_dynsyst}) 
as
$x(\tau+1)=h(x(\tau))$ and $dx/d\tau=h(x)$, 
respectively,
we can write
$$
K^*=
-\frac{1}{k!}
\frac{d^{k} h(p^*)}{d x^k}
$$
(with $k\ge 2$ the lowest 
order of the derivative 
for which $h^{(k)} (p)$ is not zero)
and 
\begin{equation}
y=\tau(\mathcal{M}-\mathcal{M}^*),
\label{ytauMM}
\end{equation}
with 
$$\mathcal{M}=h'(p)$$
playing the role of the (alternative) natural bifurcation parameter,
with bifurcation point given by
$\mathcal{M}^*=h'(p^*)$
(equal to 1 in the discrete case and to 0 in the continuous case).
%
%
As always, the scaling function $G$ is given by Eq. (\ref{scalingfunc}).




\section{Scaling law for the distance to the attractor}

\subsection{Finite-time scaling for the approach to the attractor}

We have already applied the scaling laws,
unified in Eq. (\ref{lleipm}),
to the case in which 
the fixed point $p$ 
can be stable or unstable, 
corresponding to a change in stability when crossing a bifurcation,
with the only requirement, in practice, 
that $p$ is constant 
(in order that the derivatives of $h(x)$ up to order $k-1$ are zero at $p$, excluding order one).
As at the bifurcation point the fixed point $p$ changes from stable to unstable (or vice versa),
the resulting $x(\tau)-p$ changes from nearly zero to a different value,
as seen in the previous figures.
This holds, for instance, for the bifurcations given by the $f(x)$ in Eq. (\ref{fxk}) 
and their continuous counterpart
(transcritical, pitchfork...), but not for the saddle-node bifurcation.
 
However, in addition, the scaling law is also valid when $p$
is replaced by a stable fixed point $q$
(departing $p$ when this becomes unstable).
In such a case, $x(\tau)-q$ 
quantifies the ``distance'' to the attractor and
has 
always
to approach zero; 
this case is contained also in the scaling law, Eq. (\ref{lleipm}).
But, despite of the fact that the scaling law is the same, the conditions for its validity
are different. 
We will have two main subcases.
Either $q$ is constant, as given, e.g., by Eq. (\ref{fxk}) for $\mu<0$,
where the system will be at one side of the bifurcation
(this constitutes, obviously, a subset of the previous case given by $p$),
or, in the second possibility, $k=2$,
and $q$ will represent two different stable fixed points, 
one at each side of the bifurcation
(in the case in which the bifurcation has stable fixed points at both sides).
The discrete version of this was studied in Ref. \cite{Corral_Alseda_Sardanyes}.

In this second subcase the unified scaling law Eq. (\ref{lleipm}) simplifies to
%
%
\begin{equation}
x(\tau)-q=\frac 1 {K_2^*\tau}\,\tilde G(z),
\label{casok2}
\end{equation}
with $z=\tau (h'(q)-h'(q^*)) \le 0$,
from Eq. (\ref{ytauMM}),
and $K_2^*=-h''(q^*)/2$.
The subscript 2 indicates that 
the result is only valid for $k=2$,
i.e., for the transcritical and the saddle-node bifurcations.
The ``new'' scaling function is given by
$$
\tilde G(z) =\frac{|z|}{e^{|z|}-1},
$$
and verifies in fact $\tilde G(z)=G(z)$ for $z\le 0$
($z$ is defined to be in that range, where $z=y$),
but $\tilde G(y)\ne G(y)$ when $y>0$.
The different notation for this scaling function is in order to stress this.

In summary, when the stable fixed point $q$ is constant
(at one side of the bifurcation), the scaling law (\ref{lleipm})
is perfectly valid, just replacing $p$ by $q$;
and in order to stress that $y$ can only take negative values, 
we may replace $G(y)$ by $\tilde G(z)$.
This holds for the example defined by Eq. (\ref{fxk}), 
but not for the saddle-node bifurcation.
In contrast, 
when $k=2$ (transcritical and saddle-node bifurcations)
we have that the scaling law (\ref{lleipm}) transforms into a simplified form,
given by
Eq. (\ref{casok2}).
Nevertheless, for the transcritical bifurcation both subcases are the same.

\begin{figure}[h]
   \includegraphics[width=0.5\linewidth]{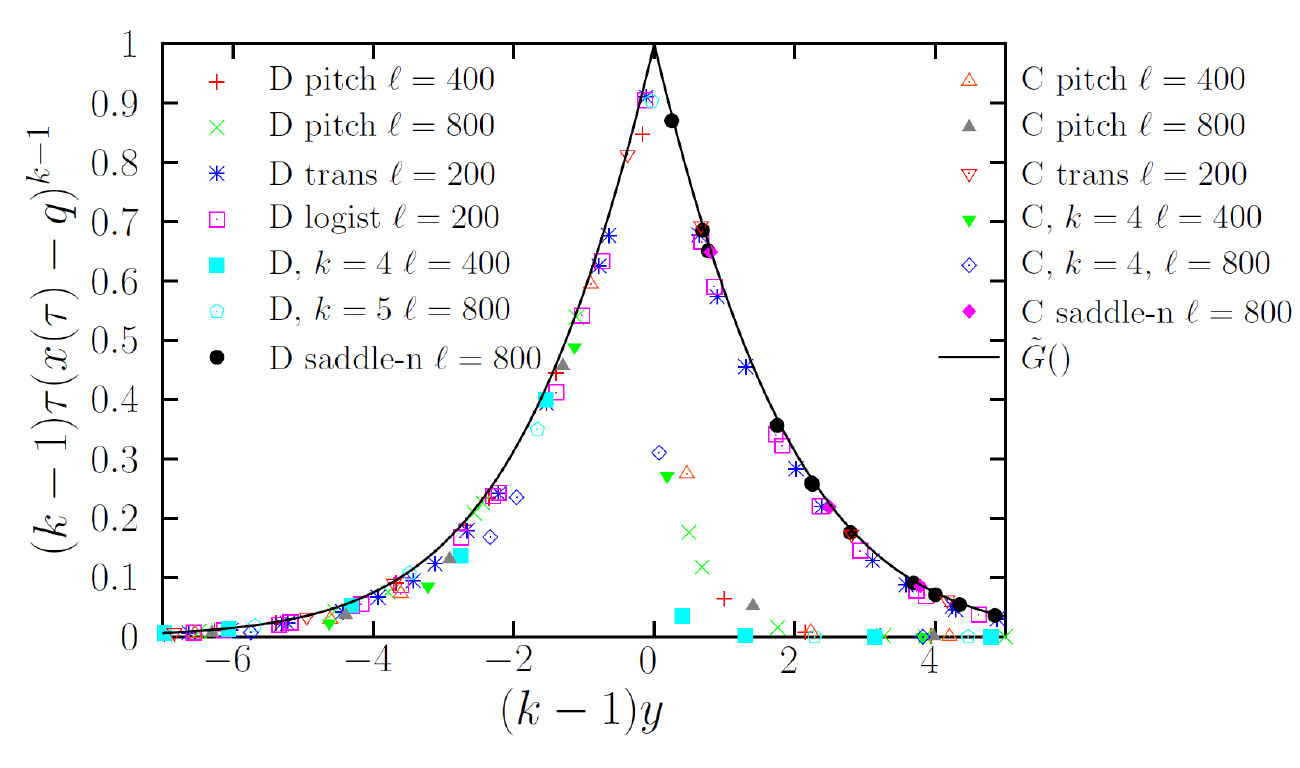}
    \caption{
Distance to the attractor, rescaled, 
versus distance of the natural bifurcation parameter to the bifurcation point, also rescaled, 
at different values of the time
for the discrete (D) dynamical systems contained in Eq. (\ref{fxk}),
for the saddle-node bifurcation, given by
$f(x)=\mu+x-x^2$, 
and for their continuous (C) counterparts.
In concrete, the horizontal axis displays $(k-1)y$,
with $y=\tau(\mu-\mu^*)$ for all cases except for the saddle node, for which
$y=-2\tau \sqrt{\mu}$ 
(we have omited this minus sign for the sake of visualization).
$\mu^*=0$ in all cases, except for the logistic one.
Initial conditions as in previous figures.
}
    \label{figure4}
\end{figure}

On the other hand,
when $q$ is not constant and $k\ge 3$,
the scaling law (\ref{casok2}) does not apply
but
one can straightforwardly compute the distance to the attractor if one is able to 
calculate the value of the attractor (as a function of the bifurcation parameter),
using that the scaling law is not valid for the attractor $q$ but for another fixed point $p$
if this keeps a constant value.
Obviously, $x(\tau)-q=x(\tau)-p + p-q$, where one just needs to substitute 
the scaling law Eq. (\ref{lleipm}) and the dependence of $q$ with $\mu$ (and therefore with $\mathcal{M}$).
%
%
%
In the particular cases given by Eq. (\ref{fxk}), with $k\ge 2$,
and extending them also to continuous dynamical systems,
one finds that
$q=0$ for $\mu\le 0$
and $q=(\pm 1)^k \sqrt[k-1]{\mu}$ for $\mu\ge 0$,
leading to
\begin{equation}
x(\tau)-q=\frac {(\pm 1)^k} {\sqrt[k-1]{(k-1)\tau\,}}\,
\left[\sqrt[k-1]{G((k-1)\tau\mu)}-\sqrt[k-1]{(k-1)\tau\mu\,}\,\Theta((k-1)\tau\mu)\right],
\label{lalarga}
\end{equation}
when $x(0)$ is above the unique attractor $q$ for $k$ even
and $|x(0)|$ is above the absolute value $|q|$ of the two attractors for $k$ odd,
with $\Theta()$ being the Heaviside (unit step) function.
Note that the previous equation constitutes also a scaling law, 
and it only agrees with Ref. \cite{Corral_Alseda_Sardanyes}
for $k=2$, for which $z=-|y|$
and we recover Eq. (\ref{casok2}).
This is illustrated in Fig. \ref{figure4}, where the shape of the scaling function $\tilde G(y)$
emerges from the data collapse in the cases in which the conditions for the validity of the scaling law
are fulfilled (in the opposite case, $k\ge 3$ and $\mu>0$, one can observe how the scaling law is not fulfilled).
Table \ref{table2} provides a summary of the rescaling necessary for each variable in order that the scaling law is fulfilled.

%

\begin{table}[h]
\begin{center}
\caption{\label{table2}
Rescaled axes for $\mu$ (horizontal axis) and $(x-p)$ (vertical axis)
in order that the scaling laws hold.
The two columns on the left refer to particular rescalings for each bifurcation, 
in which the resulting scaling functions depend on $k$
(tipically as a power of $G()$ with exponent $1/(k-1)$).
The two columns on the right lead to a universal scaling for different bifurcations.
For the case of Eq. (\ref{fxk}) with $k\ge 3$, and its continuous counterparts,
the rescaling on the right leads to the scaling function $G()$ when 
the fixed point is constant, denoted here as $p_{ct}$
(if an attractive fixed point $q$ is used instead, 
a scaling law still holds, but with different scaling functions, see Eq. (\ref{lalarga})).
For the saddle-node bifurcation, given by 
$f(x)=\mu+x-x^2$ or $g(x)=\mu-x^2$,
the scaling law holds for the distance to the attractor $q$,
yielding $\tilde G$.
The transcritical and the logistic (which has also a transcritical bifurcation)
fulfill both scaling laws.
}
\smallskip             
\begin{tabular}{|c|cr|cr|}
\hline
bifurcation
&horizontal & vertical \phantom{....} & horizontal & vertical \phantom{.....}\\
\hline
trans
& $\tau \mu$
& $\tau(x-p)$
& $\tau \mu$
& $\tau(x-p)\phantom{^{k-1}}$\\
logistic
 &$\tau (\mu-1)$
& $\tau(x-p)$
& $\tau (\mu-1)$
& $\tau(x-p)\phantom{^{k-1}}$\\
pitchfork
&$\tau \mu$
& $\sqrt{2\tau}\,(x-p_{ct})$
& $2\tau \mu$
& $2\tau(x-p_{ct})^2\phantom{^{-1}}$\\
Eq. (\ref{fxk})
& $\tau \mu$
& $\sqrt[k-1]{(k-1)\tau}\,(x-p_{ct})$
&$(k-1)\tau \mu$ \,
& \, $(k-1)\tau(x-p_{ct})^{k-1}$\\
saddle-n
&$4 \tau^2 \mu$
& $\tau(x-q)$
&$2\tau\sqrt{\mu}$
& $\tau(x-q)\phantom{^{k-1}}$\\
Eq. (\ref{mapraiz})
&$4 \tau^2 \mu$
& $\tau(x-q)$
&$\pm 2\tau\sqrt{|\mu|}$
& $\tau(x-q)\phantom{^{k-1}}$\\
\hline
\end{tabular}
\par
\end{center}
\end{table}

\subsection{Illustration of validity of the scaling law beyond its limits of validity}

We now analyze the approach to the attractor 
in a case in which there is not a strict bifurcation.
Let us consider a discrete dynamical system given by
\begin{equation}
f(x)=\mbox{sign}(x+\mu) \left(\frac{\sqrt{1+4|x+\mu|}-1}{2}\right),
\label{mapraiz}
\end{equation}
with $\mbox{sign}()$ giving the sign of its argument (being zero when this is zero).
This apparently complicated map arises as the inverse of the very simple map
$f(x)=-\mu+x+\mbox{sign}(x) x^2$, 
where all the fixed points, for different $\mu$,
except $\mu=0$, are linearly unstable
(the reason for writing $\mbox{sign}(x) x^2$ and not $x^3$ in $f(x)$ 
is in order to facilitate its inversion).
Therefore, in the case of the map given by Eq. (\ref{mapraiz}),
the ``bifurcation diagram'' consists of just one linearly stable fixed point (except for $\mu=0$).
This is given by
$p=q=-\sqrt{-\mu}$ for $\mu < 0$
and
$p=q=\sqrt{\mu}$ for $\mu > 0$,
(and $p=0$, which is marginal for $\mu=\mu^*=0$).
Thus, nothing bifurcates at $\mu=0$ but still we have a qualitative different behavior
just at that single point.

\begin{figure}[h]
   \includegraphics[width=0.5\linewidth]{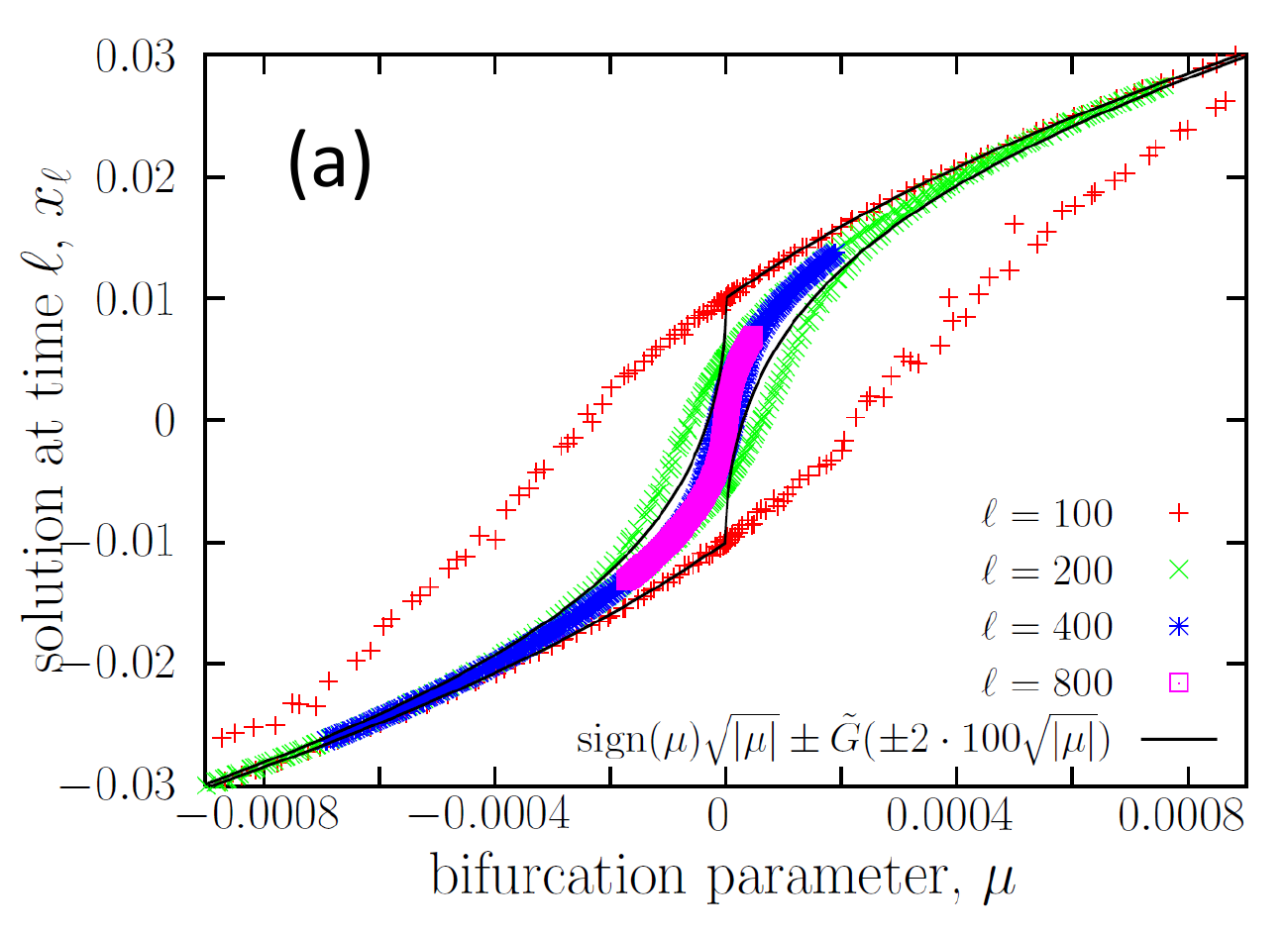}
   \includegraphics[width=0.5\linewidth]{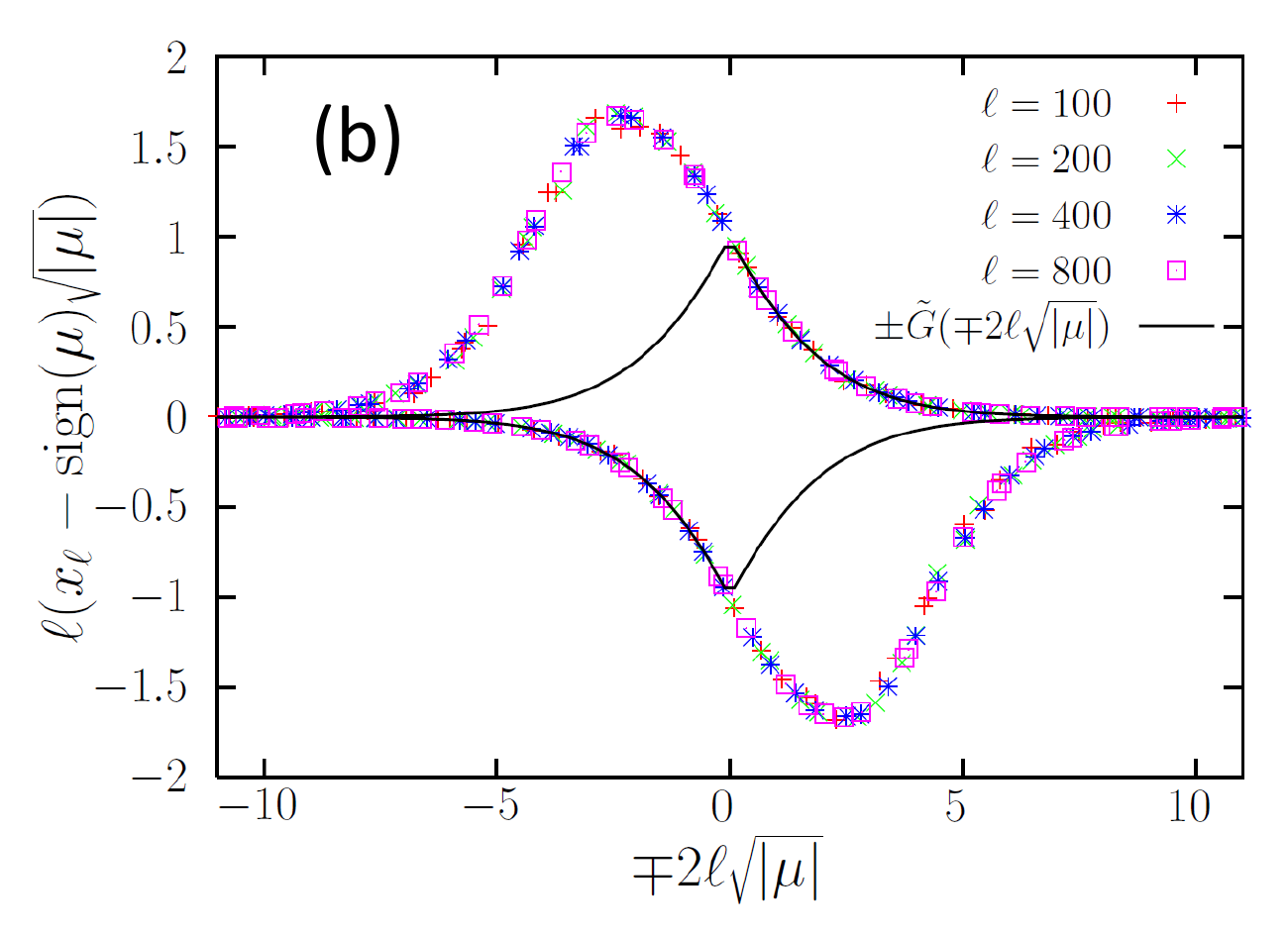}
    \caption{
(a) $x_\ell$ versus $\mu$ at different values of time $\ell$ for the 
discrete dynamical system given by Eq. (\ref{mapraiz}).
For increasing times, the solution $x_\ell$ approaches 
$q=\mbox{sign}(\mu)\sqrt{|\mu|}$
(b) Distance to the attractor 
with rescaled axes (reducing the number of data points, for clarity sake).
For the sake of visualization, 
we change the sign of $y$ when $\mu>0$, 
i.e.,
$y= 2\tau\sqrt{\mu}$ in this case.
A finite-time scaling law is fulfilled, but the scaling function is only valid (as expected) for a half of the parameter space.
The maximum distance is attained at about 2.4 in the horizontal axis.
}
    \label{figure5}
\end{figure}

We test the scaling law for such a dynamical system, 
for which $y\simeq -2\tau\sqrt{|\mu|}$ 
(for $\mu$ small)
and $K=\mbox{sign}(\mu)$
(except for $\mu=0$, for which the second derivative at the fixed point is not defined, 
and thus there is not a value of $K^*$ that we can use).
Figure \ref{figure5}(a) displays the solution $x(\tau)$ for different values of $\mu$ at several times $\tau$,
showing how $x(\tau)$ approaches the attractor $q$ for different values of $\mu$ as $\tau$ increases.
Figure \ref{figure5}(b) represents the corresponding rescaled distance to the attractor as dictated by 
Eq. (\ref{casok2}) (remember, a particular case of 
Eq. (\ref{lleipm}), but in this case without substituting
$K$ by $K^*$).
We obtain a very good data collapse, 
meaning that a scaling law holds, 
but notice that the scaling function $\tilde G(z)$ does not always describe the rescaled data.

In fact, the scaling function only works
for the cases in which the validity of the scaling law is fulfilled, 
which corresponds to when $x(\tau)-q$ and $K$ have the same sign, 
that is,
$x(\tau)>q$ when $\mu>0$
and
$x(\tau)<q$ when $\mu<0$.
This validity demonstrates that 
$\mu=0$ is somehow a critical point, 
despite one cannot consider that there is a bifurcation there.
It is noteworthy that a scaling law is still fulfilled 
in the cases in which our derivation is not valid
(corresponding to a range of 
$\mu$ and $x(\tau)$ 
opposite to the one given above), 
and in particular, it is intriguing that the maximum distance to the attractor $q$ is not attained at the critical point $\mu=0$,
as described by $\tilde G(z)$, see Fig. \ref{figure5}(b).
From this figure, we can numerically obtain 
the value of $\mu$ for which the convergence to the attractor is slower, which is
given by
$2\tau \sqrt{|\mu_\text{slow}|}\simeq 2.4$,
and thus,
$\mu_\text{slow}\simeq \pm 1.4/\tau^2$.
In summary, we have explained one situation in which the scaling law is valid but the scaling function is not
(but still it is valid for a limited range of $\mu$, as dictated by our analytical calculations).

%



\section{Conclusions}

We have shown how a very important concept in statistical physics, finite-size scaling, can be used to illuminate the finite-time dynamics in dynamical systems
close to a bifurcation point.
Our results extend the previous ones from Ref. \cite{Corral_Alseda_Sardanyes} in two directions: beyond the transcritical and saddle-node bifurcations and beyond discrete dynamical systems to the continuous (differential equations) case.
Notably, in contrast to Ref. \cite{Corral_Alseda_Sardanyes}, our approach does not need to calculate the value of a stable fixed point, and can proceed with the knowledge of a unstable one (which in many cases can be simpler to know).
The key result is the finite-time scaling law, Eq. (\ref{lleipm}), together with the definition of the rescaled distance $y$ to the bifurcation point, Eq. (\ref{ytauMM}).
It is striking the universality of the scaling function $G(y)$, Eq. (\ref{scalingfunc}),
which arises (raised to some power if $k\ge 3$) in all finite-time scaling laws that we obtain.

The parallelisms between bifurcations in dynamical systems and phase transitions in statistical physics are remarkable, and are summarized in Table \ref{table1}.
In this way, we conclude that the sharp change of behavior that characterizes bifurcations
at their bifurcation point at infinite time
arises in the same way as phase transitions arise in the infinite system-size limit
(known as the thermodynamic limit).
The way in which both asymptotic behaviors arise is through a scaling law, 
with a smooth scaling function for which the infinite limit of one of its variables
(time or system size) brings the asymptotic behavior of the scaling function to the transition point
(bifurcation point or critical point).
Then,
the ``critical'' exponents describing the transition are obviously contained in the scaling law
and in the scaling function.






\section{Acknowledgements}

We are grateful to 
Aina Cornell\`a, who participated in the early steps of this work, 
and to
J. Sardany\'es, for discussions and a critical reading.
The Centre de Recerca Matem\`atica is supported by the CERCA Programme of the Generalitat de Catalunya as well as by the Spanish State Research Agency (AEI) through the Severo Ochoa and Mar\'{\i}a de Maeztu Program for Centers and Units of Excellence in R\&D (CEX2020-001084-M).
The research of the author is supported by the projects PGC-FIS2018-099629-B-I00
and PID2021-125979OB-I00, AEI.

\section{Appendix: Critical slowing down}
\label{section_slow}

\subsection{Power-law decay at the bifurcation point} 


Naturally, the previous 
finite-time scaling law
(which has two ``versions'',
one for the bifurcation profile and another for the distance to the attractor)
include the case in which the system is right at the bifurcation point, 
$y=0$ 
(i.e., $\mathcal{M}=\mathcal{M}^*$).
As $G(0)=1$,
we obtain from Eq. (\ref{lleipm}) a power-law decay with $\tau$ towards the attractor, given by
\begin{equation}
x^*(\tau)-q^*=
\frac {(\pm 1)^k} {\left[{(k-1) K^* \tau}\right]^{1/(k-1)}},
\label{criticalslowingdown}
\end{equation}
which is in fact the generalization 
(including both discrete and continuous systems)
of one of the cases 
in Eq. (\ref{ladelallave}),
where we have replaced $p$ by $q^*$ as the decay indicates that 
the fixed point at $\mathcal{M}^*$,
despite not being linearly stable,
is indeed an attractor.
The asterisk in $x^*(\tau)$ stresses that we are right at the bifurcation in the parameter space,
and the exponent $1/(k-1)$ quantifies that the higher the value of $k$, the slower the power-law decay.
This equation is valid for the transcritical and pitchfork bifurcations, 
as represented by Eq. (\ref{fxk}), including higher values of $k$, 
as well as for their continuous counterparts, 
and also for the saddle-node bifurcation, replacing $p$ by $q$ in Eq. (\ref{lleipm}).
In concrete, 
$x^*(\tau)-q^*\propto 1/\tau$
for the transcritical and saddle-node bifurcations, 
$|x^*(\tau)-q^*|\propto 1/\sqrt\tau$
for the pitchfork case,
$x^*(\tau)-q^*\propto 1/\sqrt[3]{\tau}$ for $k=4$ and so on.
A power-law decay in time when the system is in a critical point is referred to in physics as 
critical slowing down \cite{Strogatz_book}.
The fulfillment of the scaling in the particular case $y=0$ in
Figs. \ref{pitch}(b), \ref{k4}(b)-(c), \ref{Figcont}(b), \ref{figure4}, and \ref{figure5}(b)
provides visual verification of this critical slowing down.

\subsection{Divergence of the characteristic decay time}

The scaling law (\ref{lleipm}),
through the rescaled variable $y$, Eq. (\ref{ytauMM}), 
relates values of 
the time $\tau$, 
the bifurcation parameter $\mu$ (or $\mathcal{M}$),
and 
the rescaled distance of the solution to the fixed point
that are ``corresponding'', in the sense that they yield the same value of the function $G()$,
and thus, these ``corresponding'' parameters and values
are uniquely described by the value of $G()$.
But the law can also be used to compare the dynamics at fixed $\mu$ for different $\tau$.
Outside the bifurcation point, we consider first with the case $y < 0$.
Then, 
$G(u) =\tilde G(u) \rightarrow u e^u$ 
when 
$u=(k-1)y=(k-1) \tau (\mathcal{M}-\mathcal{M}^*) \rightarrow -\infty$,
and substituting in Eq. (\ref{lleipm}) we obtain
\begin{equation}
x(\tau)-q = (\pm 1)^k \sqrt[k-1]
{\frac{|\mathcal{M}-\mathcal{M}^*|}{K^*}} \, e^{-\tau / \tau_c}\,
\mbox{ for } \mathcal{M}<\mathcal{M}^*,
\label{ladearriba}
\end{equation}
thus, the decay is exponential,
with $\tau_c$ a characteristic decay time given by
$$
\tau_c=\frac 1 {\mathcal{M}^*-\mathcal{M}}.
$$
In the opposite case, $y>0$, we use Eq. (\ref{lalarga}), 
which describes the results for the dynamical system given by Eq. (\ref{fxk}).
When $u \rightarrow \infty$ we can substitute 
$$
\sqrt[k-1]{G(u)} - \sqrt[k-1]{u} \rightarrow \frac 1{k-1} {\sqrt[k-1]{u}\, e^{-u}},
$$
to arrive to Eq. (\ref{ladearriba}) with $K^*=1$ but
$$
\tau_c=\frac 1 {(k-1)(\mathcal{M}-\mathcal{M}^*)}
\mbox{ for } \mathcal{M}>\mathcal{M}^*.
$$
In this way, the characteristic time $\tau_c$ is different at each side of the bifurcation
if $k\ge 3$, and this different behavior is fully contained in the scaling function.
In any case, notice that at the bifurcation point, $\tau_c\rightarrow \infty$ 
and the exponential decay is no longer valid.
Instead, Eq. (\ref{criticalslowingdown}) applies.

We have shown that 
$\tau_c$ decays hyperbolically
in terms of the natural bifurcation parameter $\mathcal{M}$.
For the transcritical bifurcation, the pitchfork bifurcations,
and the cases given by higher values of $k$ arising from Eq. (\ref{fxk}),
we have that
$\mathcal{M}-\mathcal{M}^*=\mu$
(or equal to $\mu-\mu^*$ with $\mu^*=1$ for the transcritical bifurcation contained in the logistic map);
so, the decay is the same in terms of $\mu$.
However, for the saddle-node bifurcation
$y=-z=\tau(\mathcal{M}-\mathcal{M}^*)=2\tau \sqrt{\mu}$, 
and therefore $\tau_c=1/\sqrt{4\mu}$ in this case.
Thus, the critical exponent that relates $\tau_c$ with $\mu$ 
is $1/2$ in the saddle-node bifurcation, 
but turns out to be 1 if one uses the natural bifurcation parameter $\mathcal{M}$.
The map given by Eq. (\ref{mapraiz}) displays the same behavior.






\begin{thebibliography}{10}
\bibitem{footnote_aina}
If one wants to keep track of the sign of $d_n$, defining, for instance, $c_n = \mbox{sign}(d_n)/d_n^\alpha$
for $\alpha$
even, the final results are the same.
\end{thebibliography}

\end{document}